# Harnessing Transparent Learning Analytics for Individualized Support through Auto-detection of Engagement in Face-to-Face Collaborative Learning

Qi Zhou
UCL Knowledge Lab, University College London
qtnvqz3@ucl.ac.uk

Wannapon Suraworachet
UCL Knowledge Lab, University College London
wannapon.suraworachet.20@ucl.ac.uk

Mutlu Cukurova
UCL Knowledge Lab, University College London
m.cukurova@ucl.ac.uk

## ABSTRACT

Using learning analytics to investigate and support collaborative learning has been explored for many years. Recently, automated approaches with various artificial intelligence approaches have provided promising results for modelling and predicting student engagement and performance in collaborative learning tasks. However, due to the lack of transparency and interpretability caused by the use of "black box" approaches in learning analytics design and implementation, guidance for teaching and learning practice may become a challenge. On the one hand, the black box created by machine learning algorithms and models prevents users from obtaining educationally meaningful learning and teaching suggestions. On the other hand, focusing on group and cohort level analysis *only* can make it difficult to provide specific support for individual students working in collaborative groups. This paper proposes a transparent approach to automatically detect student's individual engagement in the process of collaboration. The results show that the proposed approach can reflect student's individual engagement and can be used as an indicator to distinguish students with different collaborative learning challenges (cognitive, behavioural and emotional) and learning outcomes. The potential of the proposed collaboration analytics approach for scaffolding collaborative learning practice in face-to-face contexts is discussed and future research suggestions are provided.

**ACM Reference Format:**
Qi Zhou, Wannapon Suraworachet, and Mutlu Cukurova. 2024. Harnessing Transparent Learning Analytics for Individualized Support through Auto-detection of Engagement in Face-to-Face Collaborative Learning. In *The 14th Learning Analytics and Knowledge Conference (LAK '24), March 18–22, 2024, Kyoto, Japan.* ACM, New York, NY, USA, 12 pages. https://doi.org/10.1145/3636555.3636894

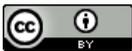

## 1 INTRODUCTION

Collaboration has been considered an important skill both in learning and working [24]. However, only asking students to work together cannot guarantee the success of collaborative learning. It requires students to have mutual interactions and achieve consensus on what they have discussed. Yet, understanding such a complex process of collaborative learning is a challenge for both teachers and students [20].

Recently, learning analytics and AI approaches have been frequently used in the context of collaborative learning. It has been argued to have the potential to contribute to investigating different modes of collaboration [35], modelling learner behaviours [28], predicting collaboration performance [17], providing adaptive support for collaborative groups [23], and theory building [27]. However, previous studies focused more on sensing and capturing data, automation of metrics generation, and elucidations of the analysis result [8]. How to design learning analytics to provide learning suggestions and feedback in collaborative learning, especially physical collaborative learning contexts, is still under exploration.

The transparency of the learning analytics systems might be a significant challenge when implementing them in natural teaching and learning scenarios [12]. To be more specific, although the non-transparent approaches, such as neural networks, can provide models with high accuracy [22], their value can be limited in education contexts. The interpretation of which factors influence the outputs of models and their metrics has been illustrated to be an important purpose of learning analytics [9]. Moreover, it is also difficult for users to understand the metrics used in such non-transparent approaches and inducted learning suggestions from their relationship with learning outcomes. For instance, previous research showed that the distance between students' hands might be a reliable predictor of group performance in hands-on, practice-based collaborative learning activities [40]. However, teachers cannot simply ask students to put their hands closer to each other to provide support for their collaboration. Therefore, the value of such a model for a practitioner is limited. There is a need to take both the transparency and the educational value of what is made transparent into consideration when designing learning analytics systems.

Therefore, this study proposes a transparent approach to automatically detecting students' individual engagement from their group interactions, aiming at providing educationally meaningful insights for teachers and learners to support individuals in a real-world face-to-face collaborative learning context.

## 2 BACKGROUND
### 2.1 Analytics of Collaborative Learning

As a research area which focuses on measuring, collecting, analysing and reporting the data about learning and learning environment [39], learning analytics has been frequently applied to collaborative learning in the past decade. It has been argued to have various benefits in providing more comprehensive information to gain a better understanding as well as fostering collaborative learning [8].

First, learning analytics has been used to automate the detection of students' behaviours. Some studies tried to detect behaviours which related to the learning contexts. For example, Martinez-Maldonado, et al., [30] presented an approach to detecting students' learning phases (e.g. instruction, simulation, and reflection) by analysing their mobility data in a collaborative healthcare simulation classroom. Whereas, some other studies focused on the detection of more general behaviours. For instance, previous studies explored how learning analytics can be used to detect gaze behaviours [48] and speaking behaviours [26]. Yet, the value to the end users of these studies might be limited because of the gap between the detected behaviours and effective feedback or learning suggestions for learners and teachers. There are various ongoing research studies to make such analytics more meaningful for end users [14]. Second, learning analytics was considered to have the potential to predict students' performance in collaborative learning. For example, Spikol and colleagues [41] proposed a supervised machine-learning method to predict students' collaborative problem-solving competencies in physical computing activities. They collected and analysed video data, audio data, and log data, and achieved the accuracy with a mean squared error of 0.13. However, the generalizability [7] and pedagogical value [31] of these predictions are also considered limited. Third, existing research illustrated the possibility of applying learning analytics to provide learning support and feedback for learners [41] and teachers [33] in collaborative learning. Specifically, Zhou et al., [52] explored the potential of a learning analytics tool, ZoomSense, which can visualise students' speaking time and turn-talking behaviours, to support students in online synchronous collaborative learning. It is reported that the learning analytics tool can help students to be aware of each other's contributions to the group discussion and make changes accordingly to foster collaboration. Nevertheless, it was found that the lack of content analysis of the discussions makes students easily game the system. Similarly, Pozdniakov and colleagues explored how this tool can benefit teachers with the help of a data storytelling approach. It is found that although it can help teachers with their sensemaking process of students' learning process, teachers were concerned about the lack of transparency in terms of the visual interface and the algorithms behind it. The authors argued that this concern reflected a wider debate on how to provide insights into the design decisions of the analytics intended to represent the end users [32].

Although much research has claimed the benefits of applying learning analytics in collaborative learning, the implementation of learning analytics systems in real-world collaborative learning settings is still facing significant challenges. For instance, given the fact that most existing learning analytics designs adopted complex data collection settings which lead to both high financial and high technical burdens [46]. Also, since the nontransparent modelling approaches were commonly used in this area, it is argued to be difficult to provide educational meaningful insights about the collaborative learning process considering the low-explainability and relevance of such approaches to educational stakeholders [12]. Moreover, because of the large amount and various modalities of data collected and analysed, ethical and privacy concerns need to be taken into account in the implementation of learning analytics designs [1]. Thus, most of the existing studies were conducted in a laboratory setting or in well-structured online collaborative learning contexts [8]. More research is needed to gain a deeper understanding of the use of learning analytics in real-world physical collaborative learning contexts.

### 2.2 Transparency of Learning Analytics Systems

The transparency of learning analytics systems has been considered to be a challenge for its adoption in practice [2, 12]. Some studies applied "black-box" approaches, which only presented the input and output of the systems, but did not provide an explanation about how the decision/judgement about learning outcomes was made. Compared with other fields in which the accurate prediction performance might be sufficient (e.g. finance, safety); learning analytics design not only requires an accurate prediction but also aims at understanding how related factors may affect a learning outcome measurement [9]. Since the non-transparent design of the learning analytics system cannot provide such explanations, it is hard for teachers or students to use the provided information to improve their teaching and learning.

Transparency of an LA system can be damaged at multiple levels in the pipeline from conceptualization of LA to its final scaled adoption in practice [6, 43]. Here we are only focusing on 1) the transparency of the models and 2) the level of analysis for an individual learner. On the one hand, the "black box" was created by the application of some recent machine learning models. Even though these models are frequently used due to their high accuracy in generating insights from complex multimodal data [22], they can hardly enable users to interrogate the analytics suggestions and provide learning support accordingly. To address this problem, Cukurova and colleagues [12] presented a transparent approach, using a decision tree, to model collaborative problem-solving competence through video data. Compared with previous studies which directly predicted learning outcomes from the computer-derived features, this study introduced different types of learners' behaviours between features and outcomes measurements. Furthermore, a similar framework, "from clicks to constructs" was raised by Wise et al., [44] and further developed by Martinez-Maldonado et al., [29] in collaboration analytics. The framework divides the data analysis in collaborative learning into three parts, namely computationally detectable features, human observable behaviours, and constructs which are not directly observable. By separating these layers in analytics, the modelling will no longer rely on the machine learning algorithm only, but also can take educational theory [45], expert knowledge [4], and human-centred approaches [5] into account.

On the other hand, the trend of targeting only group-level performance predictions in collaboration analytics might lead to another opaqueness for the practice. It is argued that conducting multilevel analysis, both individual-level and group-level, is helpful in the adoption of a learning analytics system in collaborative learning [19]. Many studies only used group-level measurements, such as the quality of the collaboration productions [50], and human-observed collaboration quality [41], as their target constructs in the analysis. Given the fact that group-level analysis only provides information about the learning performance as a collaborative group, it is a challenge to provide individual learning suggestions for students in practice. Correspondingly, Zhou and Cukurova [49] further expanded on the "from clicks to constructs" framework to propose a multilevel collaboration analytics framework called "Zoom Lens". "Zoom Lens" extended the human-observable behaviours of the previous framework into three different types of behaviours, namely social signals, group interactions and individual engagement. The group interactions were detected from each individual's social signals while the individual engagement was estimated based on its social signal and the group interaction which was happening.

In this study, we applied the "Zoom Lens" framework proposed and suggested a rule-based approach to detect individual engagement from group interactions automatically. Our ultimate goal is to operationalise the theoretical framework and provide comprehensive information about individual engagement in collaborative learning to learners and teachers. The overarching research aim of this paper is to explore the effectiveness of the automatic and transparent approach to detecting students' individual engagement from group interactions. This aim was shaped into three main research questions:

- RQ1) How does students' automatically detected engagement align with their self-reported engagement in collaborative learning?
- RQ2) Do students with higher engagement experience fewer challenges compared to students with lower engagement during the process of collaborative learning?
- RQ3) Do students with higher engagement achieve better learning outcomes compared to students with lower engagement during the process of collaborative learning?

## 3 METHODOLOGY
### 3.1 Educational Context

The study was conducted in a 10-week postgraduate module in a UK university. In this module, students were allocated into groups of 4 or 5 members with mixed-gender, mixed-cultural backgrounds, varied first languages, varied working experiences, and interdisciplinary backgrounds. During the 10 weeks, they were asked to work collaboratively on an educational technology design case to address the educational challenge they came up with in the first week.

For each week, students were asked to 1) read the literature related to the weekly topic; 2) watch the pre-recorded lectures; 3) join in the online asynchrony discussion on Moodle forum; 4) participate in a two-hour face-to-face session; and 5) write a reflection about their learning experience on Google Docs. The face-to-face sessions consisted of a one-hour tutorial session and a one-hour collaborative learning session. During the collaborative learning sessions, students were asked to make full use of the learning materials and collaborate with each other to finish the tasks which were pre-set on the Miro platform. They could ask for help from tutors if they met any challenges in this process. This study focused on data collected from these sessions.

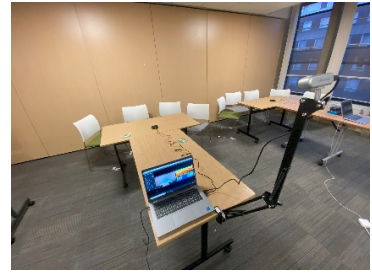

**Figure 1: Data collection setting**

The final assessment of this module consists of two parts. The first part is an individual reflection from weekly activity five, and the second part is a 3000-word essay focused on critical reflection on an existing educational technology tool. The first part occupied 40% of the final score while the second part occupied 60% of the final score. The students were marked for 20 items with grades from A to D, represented by numbers from 4 to 1 in this study. An average score (min = 0.49, max = 2.71) was calculated from these 20 items to be used as their final assessment scores in this study.

### 3.2 Data Collection

During the face-to-face collaborative learning sessions, students were asked to sit around the wide end of a T-shaped table (figure 1). Each student used their own laptop/tablet to access the online collaboration platform, Miro. A conference microphone was placed in front of them on the table to collect the audio data from their oral discussion. At the far end of the T-shaped table, an Intel RealSense was set to capture the video data.

The audio data was collected as .mp3 files with timestamps of the start time. The video data was captured as .bag files and then transferred to .mp4 files and Excel sheets with the timestamps of each video frame. The timestamps were used to synchronize the audio and video data. In this study, data from twenty collaborative sessions, lasting from about 33 minutes to 67 minutes, have been analysed.

In addition to the video and audio data, this study also collected self-reported data from students. After each session, students had to fill in a questionnaire asking about their experiences with shared-understanding building, collaboration engagement, and challenges they might meet during the collaborative learning activities. From the twenty analysed sessions, 76 valid responses were collected.

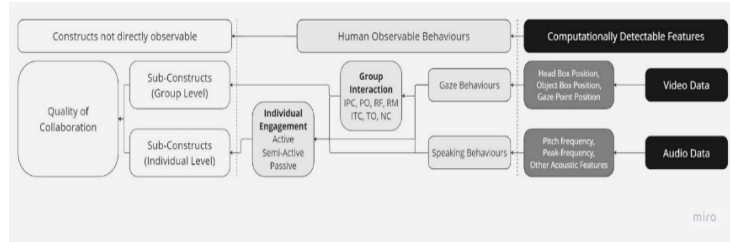

Figure 2: An overview of the data processing

## 3.3 Data Processing

Figure 2 shows an overview of the data processing. First, individual behaviours were detected from the raw data. Then, the group interactions were detected by analysing the individual behaviours of all group members. After that, the individual engagement was estimated based on each individual's behaviours and the group interactions. At last, the estimated individual engagement was connected with students' collaborative learning experiences.

*3.3.1 Detecting Individual Behaviours.* The first step is detecting the individual behaviours from the raw data. In this study, gaze behaviours and speech behaviours were detected from the video data and audio data separately.

Four types of gaze behaviours were identified in this study, namely gazing at peers, gazing at laptops/tablets, gazing at tutors and gazing at other objects. The pattern of these gaze behaviours has been illustrated to have the potential to distinguish the students who exhibited different learning processes in collaborative learning.[50] Also, these gaze behaviours can be automatically detected by computer vision algorithms with acceptable accuracy[48]. In order to work with a more accurate ground truth to answer the research questions posed, in this study, the gaze behaviours were annotated manually by two researchers. They achieved 0.98 Cohen's Kappa in a 1000-frame sample. The annotation was conducted in a second-based window.

In terms of the speaking behaviours, Speaker Diarization was conducted to identify the speakers from the audio data. To be more specific, the audio data was automatically analysed by Amazon Transcribe which is an automatic speech recognition service from Amazon Web Service (AWS). The output of the automatic tool is .json files, which provide information about the content of speaking, speaker ID, and timestamp of the start and end time of a specific speech episode. Then these .json files were converted into .csv files which showed the speaker of each second during the collaborative learning activities. Given the fact that there was a lack of pre-recorded voice samples from each student, the voice-in audio data was at this stage manually mapped to the individual students.

*3.3.2 Detecting Group Interactions.* Then, the group interactions were detected based on all group members' speech behaviours and gaze behaviours. Seven types of group interactions have been identified from students' gaze behaviours and speaking behaviours using a rule-based method presented in previous research [51] which illustrated how these types of group activities can be used to distinguish the collaborative learning process based on different learning outcomes. When we identified the group interactions, we used a slide-window approach to transfer students' gaze behaviours at a given moment into a continuous behaviour of "paying attention to". This approach can help mitigate the effects of a single disturbance in a sequence of behaviours. In this study, a five-second window was used in this step.

***Interacting with peers through communication (IPC).*** Previous research illustrated that verbal communication between group members can contribute to the building of shared understanding and socially shared regulation of collaborative learning [31]. In this study, IPC presented this type of interaction. It is used to describe a situation in which one or more than one student spoke and other students were active listening.

***Referring and following (RF).*** Besides the oral communication between group members, the discussion about specific learning materials is also important in the process of collaborative learning. Gaze following behaviour based on peers' speech has been illustrated to be an important behaviour closely related to the shaping of shared attention in collaboration [15]. RF was identified to describe this type of interaction. More specifically, RF was used when one student was speaking and other members were looking at the same objects as the speaker.

***Peer observation (PO).*** PO refers to the situation of students trying to understand what actions were taken by others by looking at them. It can reflect students' regulation dimension of monitoring[31] and is related to the success of collaborative learning [12].

***Resource Management (RM).*** RM presents the situations in which students pay attention to the same learning materials or learning tasks on their laptops/tablets. It is demonstrated that the synchrony of interactions with learning materials or learning platforms was highly related to the effectiveness of collaborative learning [13, 37].

***Interacting with a tutor through communication (ITC).*** Students' interactions with tutors are also important in collaborative learning. Tutors can provide support with domain knowledge as well as monitoring the collaborative learning process [25]. ITC presents the situation when there was a discussion happened between students and tutor(s).

***Tutor observation (TO).*** TO is another type of interaction that happens between students and tutors. It is defined to describe the situation in which students were actively listening to a tutor who might be explaining specific content, answering questions, and clarifying learning activities. TO differed from ITC since the tutors' role in this interaction was more likely to be a "lecturer"/"presenter" in a more traditional pedagogical approach [25].

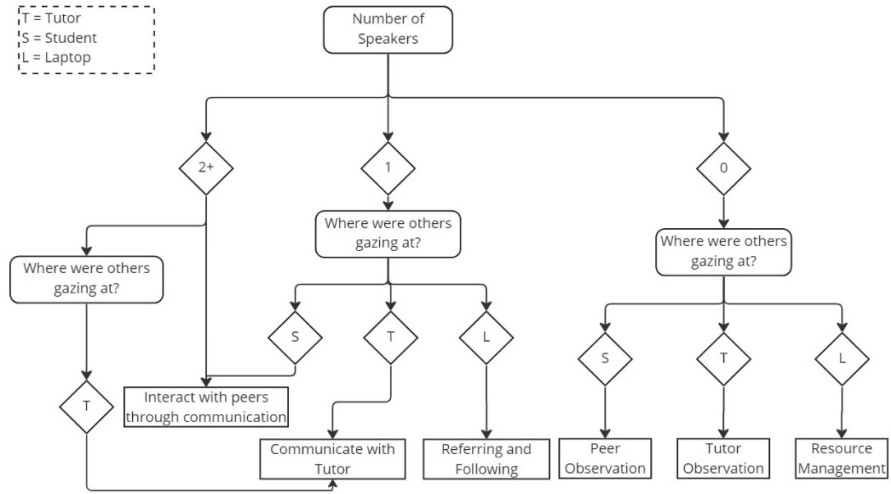

Figure 3: The rules applied for inferring group interaction statuses from the gaze and speaking behaviours

Table 1: The rules for determining the individual engagement

|     | Speech Behaviour | | | | | | | |
| --- | --- | --- | --- | --- | --- | --- | --- | --- |
|     | Speaking | | | | Not Speaking | | | |
|     | Gaze behaviours | | | | Gaze behaviours | | | |
|     | Student | Laptop | Tutor | Other | Student | Laptop | Tutor | Other |
| IPC | Active | Active | N/A | Active | Semi-active | Passive | Passive | Passive |
| RF  | N/A | Active | N/A | N/A | Passive | Semi-active | Passive | Passive |
| PO  | N/A | N/A | N/A | N/A | Semi-Active | Passive | Passive | Passive |
| RM  | N/A | N/A | N/A | N/A | Passive | Semi-active | Passive | Passive |
| ITC | Active | Active | Active | Active | Semi-active (Speaker) | Passive | Semi-active | Passive |
| TO  | N/A | N/A | N/A | N/A | Passive | Passive | Semi-Active | Passive |
| NC  | Passive | Passive | Passive | Passive | Passive | Passive | Passive | Passive |

**No collaboration (NC).** NC is used when no group interactions mentioned above were detected.

Figure 3 shows how these group interactions can be detected from each individual's gaze behaviours and speaking behaviours. where S, T and L denotes student(S), tutor(s) and laptop(s), respectively.

3.3.3 *Estimating Individual Engagement.* The third step is to estimate students' individual engagements based on their behaviours and the corresponding group interactions.

Three codes were used to present individual students' engagement in each type of group interaction introduced above, namely active, semi-active, and passive. The active code was used when a student expressed their thoughts to others or was observed by other students. The semi-active code was used when a student was actively listening to others (e.g. peers, or tutors), observing others, or paying attention to the material focused by the group. The passive code was used whenever a student was not engaged in the group interactions or the group members were not in collaboration with each other.

These three codes were generated automatically based on each individual student's speech behaviours and gaze behaviours, and the corresponding interactions they were having in the group. A rule-based method was used to determine the individual engagement of students in each second. The rules are presented in detail as table. 1. Code N/A is used if the situation is nonexistent.

For example, the first row of the figure.4 shows that this group was having a referral and follow interaction, in which one student was presenting the material on the laptop and others gazed at the same material by following the speech. Therefore, S2, the student who was speaking and gazing at the laptop was coded as "active", and others who were gazing at the laptop at the same time but not speaking were coded as "semi-active". Similarly, the second row of figure 4 explained how students were coded when they were having IPC interaction. S2 is the person who was speaking, so S2 is coded "active". S3 and S4 were gazing at students who were speaking, so they were coded as "semi-active" since they were assumed to be actively listening to S2. S1 is coded as "passive" because S1 was not participating in the group interaction.

| Session | Second | Group Interaction | Behaviours | | | | Engagement | | | |
|---|---|---|---|---|---|---|---|---|---|---|
| | | | S1 | S2 | S3 | S4 | S1 | S2 | S3 | S4 |
| W9G2 | 2 | RF | Laptop/N | Laptop/Y | Laptop/N | Laptop/N | Semi-active | Active | Semi-active | Semi-active |
| W9G2 | 314 | IPC | Other/N | Student/Y | Student/N | Student/N | Passive | Active | Semi-active | Semi-active |

Figure 4: An example of pre-processed data

At last, the frequencies of each code exhibited by each student in each session were generated.

*3.3.4 Clustering Students' Individual Engagement Tactics.* Before relating students' engagement with their self-reported data, clustering was conducted to divide them into different types according to the engagement tactics used in the collaborative learning sessions. The reasons for clustering might be various. On the one hand, gathering information from all three codes exhibited by one student might provide more information about how this student engaged in collaborative learning across the whole session. On the other hand, simply taking these codes as different digits (such as 2, 1, and 0) and calculating the cumulative measurement cannot provide meaningful enough information for both teaching and learning since it is difficult to determine the quantitative relationship between the contribution made by sharing their understanding and the contribution made by active listening.

Therefore, K-means clustering is used in this step. The normalized frequencies of all three codes from each individual student were used as the inputs. BregmanDivergences was chosen as measure types and the number of maximum runs was 10. Comparing the Avg. within the centroid distance of different numbers of clusters and considering the explainability of the clusters, two clusters (average within centroid distance = -0.026) were selected. Also, the elbow plot of the within-cluster sum of squares against the number of clusters indicates the appropriateness of k=2 with the most significant drop. There are 48 students in the first cluster and 28 students in the second cluster. Figure 5 shows the average frequencies of each code from these two clusters. The first cluster(blue) exhibited 0.316 average frequency of active code, 0.425 average frequency of semi-active code, and 0.259 average frequency of passive code. The second cluster(red) exhibited 0.091 average frequency of active code, 0.600 average frequency of semi-active code, and 0.309 average frequency of passive code. Compared with the second cluster, the first cluster exhibited more frequency of active code and less frequency of semi-active as well as passive codes.

The distinction between these two clusters is motivated by the work of Shaer, et al., [38] and Schneider, et al., [36] in which students were categorized into "drivers" and "passengers" based on their engagement in collaborative learning activities. The "drivers" were described as the students who were physically active and controlled the process of collaboration, while the "passengers" were described as the students who were physically inactive and merely proposed verbal suggestions. Since the difference between the contexts, this study mainly took students' oral expressions as their active engagement. Therefore, the first cluster was considered to be the "drivers" while the second cluster was considered to be the "passengers" of collaborative learning discourse. It is worth noting that, with improvements in sample size as well as the inclusion of more features, the number of clusters should be reconsidered accordingly in future explorations.

*3.3.5 Comparison between different engagement types using t-tests.* In order to answer the research questions, t-tests are used across the two clusters of students who take the roles of "drivers" or "passengers" during collaborative learning activities.

For the first research question, the comparison of students' self-reported engagement between two types was conducted to investigate whether the automatically detected individual engagement aligns with students' self-reports. The data used is not only students' report about their own engagement but also their report about how other group members engaged in the collaboration.

The hypotheses are:

- "Drivers" report higher engagement of their own than "passengers" in collaborative learning.
- "Drivers" report higher engagement of their peers than "passengers" in collaborative learning.

In terms of the second research question, students' self-reports about the challenges they met in the process of collaborative learning were compared across the two clusters. The report asked students to report how they felt being challenged in the aspects of behaviour, cognition, and social emotion.

The hypotheses are:

- "Drivers" faced fewer problems in behavioural/ cognitive/ socio-emotional aspects than "passengers" in collaborative learning.

In order to answer the last research question, this study also conducted a t-test between "drivers" and "passengers" in terms of the scores they achieved in the final assessment.

The hypothesis is:

- "Drivers" achieve higher grades than "passengers" in the final assessment.

All the self-reported data mentioned above was collected from a five-point Likert scale and the final grades were received based on students' assessment of the 10-week module.

## 4 RESULT

### 4.1 Comparison test on students' self-reported engagement in collaborative learning

Table 2 shows the result of the t-test of students' self-reported engagement between "drivers" and "passengers".

Both t-tests on self-reported engagement and self-reported peer engagement show significant results with p<0.05. For instance, "drivers" (n = 28, Mean = 3.86, SD = .356) had significantly higher self-reported engagement scores than "passengers" (n = 48, Mean = 3.60, SD = .707), p = .021, Cohen's d = .420. It illustrates that "drivers" are more likely to report that they were engaged in the collaboration actively than "passengers". It shows the alignments between the automatically detected individual engagement and self-reported individual engagement.

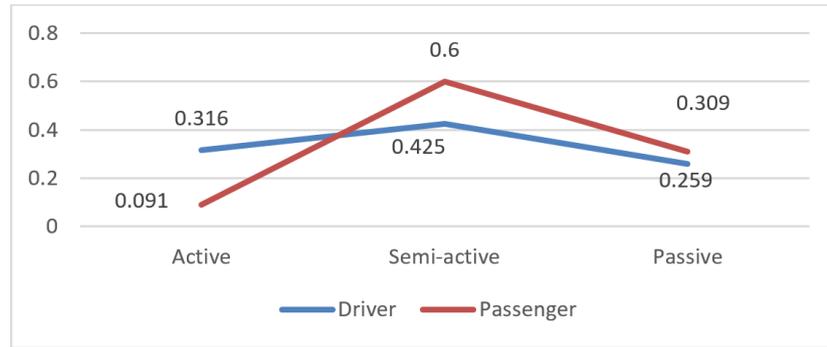

Figure 5: The average frequencies of each code from "drivers" and "passengers"

Table 2: the results of t-tests on students' self-reported engagement

|  | Drivers | | | Passengers | | | P value | Cohen's d |
|---|---|---|---|---|---|---|---|---|
|  | N | Mean | SD | N | Mean | SD | | |
| Self-reported Individual Engagement | 28 | 3.86 | .356 | 48 | 3.60 | .707 | **.021** | .420 |
| Self-reported Peer Engagement | 28 | 4.04 | .429 | 48 | 3.79 | .713 | **.033** | .391 |

Table 3: the result of the t-test on students' self-reported challenges in collaborative learning

|  | Drivers | | | Passengers | | | P value | Cohen's d |
|---|---|---|---|---|---|---|---|---|
|  | N | Mean | SD | N | Mean | SD | | |
| Behavioural Aspect | 28 | 1.18 | .476 | 48 | 2.17 | 1.506 | < .001 | .800 |
| Cognitive Aspect | 28 | 1.61 | .685 | 48 | 2.33 | 1.434 | .002 | .597 |
| Social Emotional Aspect | 28 | 1.54 | .693 | 48 | 2.38 | 1.362 | < .001 | .721 |

Moreover, the result also shows that "drivers" (n = 28, Mean = 4.04, SD = .429) had significantly higher self-reported peer engagement scores than "passengers" (n = 48, Mean = 3.79, SD = .713), p = .033, Cohen's d =.391. It shows that "drivers" are more likely to report that other members of the group were engaged in the collaboration actively than "passengers".

### 4.2 Comparison test on students' self-reported challenges in collaborative learning

Table 3 illustrates the result of the t-test between students' self-reported challenges in collaborative learning for "drivers" and "passengers".

All t-tests on students' self-reported challenge aspects show significant differences between students with two engagement types. For instance, "drivers" (Mean = 1.18, SD = .476) reported significantly higher scores in behavioural challenges than "passengers" (Mean = 2.17, SD = 1.506), p < .001, Cohen's d = .800. It means that "drivers" were less likely to report a behavioural problem during collaboration than "passengers".

Similarly, "drivers" report higher scores on cognitive challenges (Mean(e) = 1.61, SD(e) = .685, Mean(r) = 2.33, SD(r) = 1.434, p = 0.02, Cohen's d = .597) as well as socio-emotional challenges (Mean(e) = 1.54, SD(e) = .693, Mean(r) = 2.38, SD(r) = 1.362, p < .001, Cohen's d = .721) than "passengers". It means "passengers" were more likely to report cognitive challenges and social-emotional challenges in the process of collaborative learning activities.

All the results illustrate that students with different engagement types exhibited statistically significant differences in dealing with problems from these aspects. It shows the potential of using these engagement types to distinguish students with different learning processes and performance in collaborative learning.

### 4.3 Comparison test on students' learning performance

Table 4 shows the result of the t-test in students' final scores between "drivers" and "passengers". It illustrates that "drivers" (n = 25, Mean = 2.04, SD = .487) gain significantly higher scores in the final assessment than the "passengers" (n = 47, Mean = 1.81, SD = .576). It means that "drivers" may have better learning outcomes compared with "passengers".

Table 4: the result of the t-test on students' final assessment scores

|  | Drivers | | | Passengers | | | P value(one side) | Cohen's d |
|---|---|---|---|---|---|---|---|---|
|  | N | Mean | SD | N | Mean | SD |  |  |
| Final Scores | 25 | 2.04 | .487 | 47 | 1.81 | .576 | .042 | .414 |

## 5 DISCUSSION

It has been many years since multimodal learning analytics has been used in physical collaborative learning [29]. Yet, developing transparent models and designing multimodal learning analytics systems with transparency are still significant concerns for many researchers [12, 46]. First, the lack of transparency leads to a low interpretability of the commonly used modelling approaches. For instance, when using machine learning or neural network algorithms, researchers can only know the result generated by the algorithms but cannot understand how the "decisions" were made. In collaborative learning, the lack of this understanding can hardly provide educationally meaningful insights into how effective collaborative learning happens. However, many research studies still prioritise "black box" approaches due to their high performance in modelling the learners and learning process as well as predicting students' learning outcomes rather than aiming for more transparent approaches to improve the practice of collaborative learning. Second, transparency for feedback in learning analytics does not only depend on the transparency of the model. Chaudhry, et al., [6] argued that the transparency of Artificial Intelligent in Education should consider three aspects, namely data transparency, algorithmic transparency, and implementation transparency. Since transparent learning analytics aims at informing actionable feedback by making learning measurable and visible [43], transparency for instructional actions is also very important. From this perspective, only focusing on the group-level analysis in collaborative learning can be another type of black box for learners and practitioners who do not have insights into how to improve their individual practice to improve collaborative learning outcomes. Existing studies mainly focus on using multimodal learning analytics to generate models of success in collaborative groups and provide support at the group level [49]. Yet, such analysis can only provide information about the group as a cohort. It is difficult to gain information about how individual students behave in collaborative learning activities. Correspondingly, it is a challenge to understand how individual students' behaviours will affect the collaboration. Thus, the neglect of individual-level analysis might be another barrier for both teachers and students to use multimodal learning analytics systems to support their teaching and learning in real-world collaborative learning practices. Taking these two concerns of transparency in multimodal learning analytics into consideration, this study presents a rule-based approach driven by the theories of collaborative learning, to automatically detect individual student engagement from group-level interactions in a real-world collaborative learning context. The use of the rule-based approach may contribute to opening the black box of predicting and modelling while individual-level analysis presented here may show the potential of specific individual feedback for learners from their group-level collaborative learning behaviour analysis.

In order to answer RQ1: *How does the automatically detected students' individual engagement align with their self-reported engagement in collaborative learning?* This study first estimated each individual student's individual engagement (active, semi-active, and passive) based on the group interactions and corresponding individual behaviours. Then, based on the frequencies of each engagement code in one specific session, students were clustered into "drivers" and "passengers" of collaboration. The drivers of collaborative discourse exhibited a higher frequency of expressing their ideas and opinions in collaboration while the passengers had a higher frequency of actively listening to others. The comparison tests showed the alignment between automatically detected individual engagement and self-reported individual engagement. The t-tests reported a significant difference in self-reported individual engagement scores between "drivers" and "passengers". This indicates that students who expressed their ideas and opinions more frequently were likely to report they are more active in collaborative learning. This alignment is important to illustrate the validity of the proposed approach to detect student engagement in collaborative learning using automated measures of speech detection and gaze behaviours [51].

Meanwhile, it is also found that there is a significant difference in self-reported peer engagement between two types of students. In other words, students who spoke to others more frequently were more likely to report their peers were engaging actively. This result is not aligned with common sense, which would consider "drivers" to be the peers of "passengers". As a result, "passengers" would be expected to report higher peer engagement. However, previous research reported that students may have misconceptions about others' engagement during collaboration without any help [52]. According to our result, some students who occupy too much speaking time in group discussions may not be aware of others' engagement in collaborative learning. Since driver students were actively discussing with each member of the group, they tended to think everyone was actively involved in collaboration. Therefore, as previous research suggested, providing information on individual student's engagement may help them to be aware of each other's contributions to collaborative learning [52]. In this study, neither the "drivers" nor the passengers were aware of others' engagement accurately, therefore needed external scaffolding to regulate their own behaviours as well as others' behaviours. The proposed analysis has the potential to be developed as a scaffolding tool for students' socially-shared and co-regulation of learning during collaboration.

Regarding RQ2: *Whether students with higher engagement experience fewer challenges compared with students with lower engagement during the process of collaborative learning?* The comparison tests showed that there were significant differences in all behavioural, cognitive, and socio-emotional challenges between "drivers" and "passengers". To be more specific, in terms of the behavioural challenges, "drivers" were less likely to have behavioural problems during the process of collaboration. In the specific context we studied, this means that they had fewer problems with getting others to participate, finishing the work, and staying on task. One possible reason might be their speech was not only about expressing their insights about the learning content, learning materials, and learning activities. What they expressed may also be related to monitoring the process of collaboration and making adjustments accordingly. It aligns with previous research which illustrated that "drivers" are the students who make plans and manage group discussions [36]. In terms of the cognitive aspects, "passengers" tended to have more cognitive problems than "drivers". They faced more challenges in understanding the tasks as well as the competence of others. The reason behind this might be that although they spent a certain amount of time listening to others and trying to understand others' opinions, given the fact that they seldom confirmed and discussed their own understanding with others, they could hardly receive reactions from others to promote the negotiation between different understanding. Therefore, achieving consensus in cognitive aspects in the group might be a challenge for "passenger" students. Similarly, previous research argued that the maintenance of the socio-emotional aspects relied on mutual communication between group members [18]. Given the fact that "drivers" might express their feelings and help others through speech, they appeared to be less likely to take maintaining group emotion as a problem. This was reflected by the significant statistical difference in the socio-emotional challenges between "drivers" and "passengers" in this study.

Lastly, the third research question concerned whether students with high engagement achieved better learning outcomes in their assessment. The results showed that there was a significant difference in students' final scores for assessment between students who act as "drivers" and students who act as "passengers". It is well established in the literature that students' engagement does indeed affect the effectiveness and performance of a group [10]. In this study, it was found that individual engagement was related to students' individual assessment results. It is worth noting that, the assessment in this study was not only related to their collaborative learning activities, but was based on their overall achievement in the course. Students were assessed by their individual reflections and their final essays, but not only how they performed in the collaboration. The statistically significant differences between drivers' and passengers' learning gains are therefore even more impressive in this context. This may illustrate that engaging more in group activities through active communication may help students with their knowledge acquisition about the content of the module which in turn helps them achieve better grades in multiple assessments. However, the causal relationship between the two factors cannot be deemed in this explorative study and requires future research.

There are some implications of this study that are worth further discussion. First, it provides an approach for the automatic detection of individual engagement in collaborative learning as well as showing the relationship between individual engagement and students' cognitive, behavioural, and socio-emotional challenges and learning outcomes. Previous research in the field highly relied on human annotation to detect individual student engagement [10, 21], which prevented the systems from being implemented in natural collaborative learning practice automatically. The proposed method enables the automatic detection of individual engagement from audio and video data at low financial and technical costs. It may provide more opportunities for researchers to explore the relationship between individual engagement and group performance and to provide more insights into what learning support can be generated from individual engagement in real-world collaborative learning practice. In addition, detecting individual engagement based on the context of group interactions can better reflect the educational meaningful engagement in collaborative learning. Some studies considered taking active actions in online collaboration, such as posting on the group discussion [47] or in face-to-face groups such as working on the tasks [10], as actively engaging in collaborative learning. Yet, only considering data signals from the active action moments but neglecting the context where the action happened or the moments of inaction cannot fully present students' engagement in collaborative learning. Rather monitoring and support as in the proposed approach might generate more meaningful interpretations of engagement and avoid survivorship bias.

Second, the use of a rule-based approach in the proposed method enhances the human agency in learning analytics design and implementation. Most existing studies applied non-transparent approaches and essentially used these algorithms to induct one or more common behavioural patterns exhibited by students with better learning outcomes. In this sense, these non-transparent approaches use models generated from others' data to represent a path to success in collaborative learning. Educators can hardly adjust the model if there is any conflict between machine-generated insights and their own understanding and interpretations. In comparison, the rule-based approach used in this study deducts this path to success on collaborative learning theories and educators' understanding of good collaboration as well as their experience of teaching in collaborative learning. Although the accuracy of the rule-based method might not be as good as non-transparent approaches, it enables educators to define what learning behaviours are important in collaborative learning and take the agency of using (or not) the suggestions from the system regarding the students' engagement feedback. Moreover, it provides flexibility for educators to adjust the system according to different contexts. Therefore, it might lead to better generalizability and adoption in practice than non-transparent approaches. This is not to argue *per se* that one approach is better than the other one. With the purpose of using learning analytics to provide actionable feedback for learning and teaching by making learning measurable and visible rather than only optimizing the automation of analytics tools in education, the combination of the inductive nature of non-transparent approaches and the deductive nature of rule-based approaches can provide more holistic information for educators to better support collaborative learning practice.

Third, this study stresses the need to inform students about individual engagement during their collaborative learning. It is found that students struggle to judge their own and each other's engagement in collaborative learning tasks. The social translucence theory argues that students' awareness relies on the information which is visible to them, and this awareness will affect what action they will take in learning [16]. In the context of collaborative learning, if one student is not aware of others' engagement because of the lack of visible information, they can hardly regulate their behaviours to improve collaborative learning. In this study, if "drivers" assumed that their peers were actively engaged, it meant that they were less likely to be aware that there was a need to give the members who did not express their ideas the space to actively engage in discussion. Therefore, visualising individual engagement through the proposed approach might be a potential support for students to socially share regulated learning. However, it is worth noting that, a previous study reported students' different experiences of engagement compared with the machine-detected engagement [52]. This is not an attempt at retrospective correction of opinions. Our goal in visualising students' engagement is to increase their visibility to each other's engagement rather than correcting their opinions about each other's engagement.

Lastly, encouraging students to engage more is not novel for teachers in providing feedback to support students in collaborative learning. What teachers face is the complexities of organizing and providing timely support for collaborative learning in a largely sized classroom [3]. In order to foster group discussion, it is important for teachers to know what specific time is good for providing support, and which specific student needs to be motivated to engage more. This is a challenge for teachers to take care of many groups and provide such support at the same time. Therefore, the contribution of this study is not necessarily highlighting the well-established learning sciences findings such as drivers of collaborative learning discussions are likely to achieve better learning outcomes. However, we propose an approach that acts more like "eyes on the back" as an embedded tool to extend human educators' capabilities of monitoring and supporting students' engagement in collaborative learning. It has the potential to support teachers to perceive more about students' engagement during the collaboration process so teachers are able to provide feedback or learning suggestions timely, individually, and specifically. This step towards more transparency is significant, yet there are many other issues in the adoption and use of these approaches in practice that relate to ethical considerations [1], socio-cultural factors of the education system [11], as well as psycho-physiological factors of individual teachers e.g. cognitive load [34] which all require further investigations.

**Limitations and Future Research.** There are some limitations of this study that should be noted. First, there is a lack of evaluation of students' engagement and challenge moments by expert observations. This study took students' self-reports as "ground truth" in the comparison tests, but the final assessment of student learning was conducted by experts. Existing literature raised doubt about the validity and reliability of self-reporting data since participants might be biased toward themselves, and it is a challenge for them to remember the details of their collaboration process accurately. Therefore, involving evaluation from experts and triangulating the results might provide a more holistic picture of collaborative learning. Second, there is a lack of emphasis on the process of collaborative learning in the proposed approach. It is argued that students' roles and their learning tactics might change during collaborative learning activities [42]. Given the fact that this study used cumulative measurements to cluster students into "drivers" and "passengers" based on their average engagement during the whole session, the transition between different roles during the process has been neglected. Therefore, it is worth exploring the process of individual engagement by applying process mining and sequence analysis approaches. Thirdly, there is a lack of emphasis on the interactions among students in other channels. This study mainly focused on students' machine-observable gaze and speaking behaviours. However, they might also interact with each other on virtual learning platforms such as Miro and Moodle. It would be beneficial if the log data from these platforms could be involved in analysing their individual engagement. Lastly, it is not explored how the proposed approach can generate visualization or feedback for teachers or students to better support their teaching and learning. In future steps, co-design and participatory design sessions will be conducted with the students to design feedback and learning interventions driven by the insights generated from the proposed approach here.

## 6 CONCLUSION

This paper presents a novel approach to automatically detecting individual engagement from group interactions in a real-world face-to-face collaborative learning context. The proposed approach avoids the use of "black-box" approaches and suggests a rule-based method to improve the transparency of the learning analytics system design and its value to teachers and learners. The approach has the potential to distinguish the actively engaged leading students of collaborative learning activities, drivers, from those students who are passively engaged following the initiating acts and speech of others, passengers. Our results indicate that these engagement clusters are consistent with students' self-reported engagement and are highly related to the challenges they meet in the process of collaboration, as well as their learning outcomes in collaborative learning.

Our future work will focus on three aspects. First, in order to gain a more accurate picture of students' processes of collaborative learning, expert observations and discourse analysis about their engagement will be conducted to further test the validity and reliability of the results proposed in this paper. Second, qualitative evaluations will be carried out to explore the value of the feedback generated from the proposed approach to both students and teachers. Third, co-design sessions will be planned out to investigate the extent to which this analysis can be developed into a usable tool for providing learning and teaching suggestions in real-world collaborative learning contexts.

## ACKNOWLEDGMENTS

We would like to thank all UCL students and faculty in the MA EdTech programme for granting permission to collect data for this study. This work was partially supported by Amazon Web Services studentship to the first author.